\documentclass[twocolumn,pra,amsmath,amssymb]{revtex4}
\usepackage{amsfonts}
\usepackage{graphicx}
\usepackage{epic}
\usepackage{eepic}
\usepackage{color}
\usepackage{bbm}

\begin{document}

\def\tr{{\rm Tr}}
\def\ra{\rangle}
\def\la{\langle}
\def\be{\begin{equation}}
\def\ee{\end{equation}}
\def\begarr{\begin{eqnarray}}
\def\endarr{\end{eqnarray}}

\newcommand{\bjt}{\frac{\beta J}2}
\newcommand{\bjth}{\frac{3\beta J}4}
\newcommand{\bjf}{\frac{\beta J}4}
\def\ha{{\hat a}}
\def\hb{{\hat b}}
\def\hu{{\hat u}}
\def\hv{{\hat v}}
\def\hc{{\hat c}}
\def\hd{{\hat d}}
\def\no{\noindent}\def\non{\nonumber}
\def\hi{\hangindent=45pt}
\def\v{\vskip 12pt}

\newcommand{\bra}[1]{\left\langle #1 \right\vert}
\newcommand{\ket}[1]{\left\vert #1 \right\rangle}

\title{Toward a fully relativistic theory of quantum information}

\author{Christoph Adami$^{1,2}$}
\address{$^{1}$Department of Physics and Astronomy\\
$^{2}$Department of Microbiology and Molecular Genetics\\
Michigan State University, East Lansing, MI 48824}

%\date{\today}

\begin{abstract}
Information theory is a statistical theory dealing with the relative
state of detectors and physical systems. Because of this physicality of information, the
classical framework of Shannon needs to be extended to deal with
quantum detectors, perhaps moving at relativistic speeds, or even within curved
space-time. Considerable progress toward such a theory has been
achieved in the last fifteen years, while much is still not
understood. This review recapitulates some milestones along this road,
and speculates about future ones.
\end{abstract}
%\body
\maketitle
\section{Preface: From Nuclei to Quantum Information}
I am sure I am one of the more junior contributors to this volume celebrating Gerry Brown's 85th birthday, and still I've known him for 25 years. I arrived as a young graduate student at Stony Brook University in 1986, and Gerry immediately introduced me to every member of his Nuclear Theory group, ending with his postdoc Ismail Zahed. He pointed to a chair in Ismail's office, said: ``You guys talk", and left. I started to work with Ismail that day, and when he was promoted to Assistant Professor I became his first graduate student. Gerry and I only started to work together closely within the last two years of my Ph.D., and the collaboration intensified when he took me on his yearly Spring visits to the Kellogg Radiation Laboratory at the California Institute of Technology. There, I had the opportunity to meet Hans Bethe, who visited Caltech every Spring to work with Gerry. Over the following years, Hans and I became good friends and Hans's influence on my growth as a scientist would end up rivaling the influence that Gerry had on me~\cite{Adami2006}. 
In particular Hans was always very interested in my shifting interests from nuclear and high energy theory first towards quantum information theory and the foundations of quantum mechanics, and then to theoretical biology. At the same time, Gerry and Hans's collaboration on the physics of binary stars and in particular black holes continued to intrigue me.  I ended up staying at Caltech for 12 years. 

While I spend most of my time now working in biology, I still sometimes return to work in physics. People like Gerry and Hans have reinforced to me the fun that comes with attempting to understand the universe's basic principles, and when lucky enough, unravel a few of them. Perhaps it is not a coincidence that one of the striking applications of quantum relativistic information theory that I describe below is to the physics of black holes. Gerry and I discussed black holes and binary stars endlessly on walks in the mountains adjacent to Caltech, and on the phone (often on Sunday mornings) when he was back in New York.
Why did I store away article after article on black holes in the 1990s when I wasn't nearly working on the subject? I am sure it was Gerry's influence, who taught me to go after your gut instinct, and not worry if you are called crazy. I've been called crazy in many a referee's review, and I've come to realize that this usually signals that I am on to something. Thus I dedicate this article to you Gerry: there are crazy things buried in here too.  

\section{Entropy and Information: Classical Theory}
Since Shannon's historical pair of papers~\cite{Shannon1948},
information theory has changed from an engineering discipline to a
full-fledged theory within physics~\cite{Landauer1991}.  While a
considerable part of Shannon's theory deals with communication
channels and codes~\cite{CoverThomas1991}, the concepts of entropy and
information he introduced are crucial to our understanding of the
physics of measurement, and turn out to be more general than
thermodynamical entropy. Thus, information theory represents an
important part of statistical physics both at equilibrium and away from it.

In the following, I present an overview of some crucial aspects of
entropy and information in classical and quantum physics, with
extensions to the special and general theory of relativity. While not
exhaustive, the treatment is at an introductory level, with pointers
to the technical literature where appropriate.

\subsection{Entropy}
The concepts of entropy and information are the central constructs of Shannon's theory.
They quantify the ability of
observers to make {\it predictions}, in particular how well an
observer equipped with a specific measurement apparatus can make
predictions about another physical system. Shannon
entropies (also known as {\em uncertainties}) are defined for
mathematical objects called {\em random variables}. A random variable
$X$ is a mathematical object that can take on a finite number of discrete states
$x_i$, where $i=1,...,N$ with probabilities $p_i$. Now, physical
systems are not mathematical objects, nor are their states necessarily
discrete. However, if we want to quantify our uncertainty about the
state of a physical system, then in reality we only need to quantify our
uncertainty about the {\em possible outcomes of a measurement} of that
system. In other words, an observer's maximal uncertainty about a system is not
a property of the system, but rather a property of the measurement
device with which the observer is about to examine the system. For example, suppose I am armed with a measurement
device that is simply a ``presence-detector''. Then the
maximal uncertainty I have about the physical system under
consideration is 1 bit, which is the amount of {\em potential information} I can obtain about that system, given this measurement device. 

As a consequence, in information theory the entropy of a physical system is undefined if we do not
specify the device that we are going to use to reduce that entropy. A
standard example for a random variable (that is also a physical system)
is the six-sided even die. Usually, the maximal entropy attributed
to this system is $\log_2(6)$ bits. Is this all there is to know about
this system? What if we are interested not only in the face of the die that is up, but also the angle that the die has made with respect to due North?
Further, since the die is physical, it is made of molecules
and these can be in different states depending on the temperature of
the system. Are those knowable? What about the state of the atoms
making up the molecules? All these could conceivably provide labels such
that the number of states to describe the die is in reality much larger. What about the
state of the nuclei? Or the quarks and gluons inside those?

This type of thinking makes it clear that we cannot speak about
the entropy of an isolated system without reference to the
coarse-graining of states that is implied by the choice of
detector (but I will comment on the continuous variable limit of entropies below).
 So, even though detectors exist that record continuous
variables (such as, say, a mercury thermometer), each detector has a
finite resolution such that it is indeed appropriate to consider only
the {\em discrete} version of the Shannon entropy, which is given in
terms of the probabilities $p_i$ as~\footnote{From now on, I shall not
indicate the basis of the logarithm, which only serves to set the
units of entropy and information (base 2, e.g., sets the unit to a ``bit'').}
 
\be H(X) = - \sum_i^N p_i \log p_i\;.  \label{entropy}\ee

For any physical system, how are those probabilities obtained? In
principle, this can be done both by experiment and by theory. Once I
have defined the $N$ possible states of my system by choosing a
detector for it, the {\it a priori} maximal entropy is defined as 

\be
H_{\rm max}= \log N\;.  
\ee 

Experiments using my detector can now sharpen my knowledge of the
system.  By tabulating the frequency with which each of the $N$ states
appears, we can estimate the probabilities $p_i$. Note, however, that
this is a biased estimate that approaches the true entropy
Eq.(\ref{entropy}) only in the limit of an infinite number of
trials~\cite{Basharin1959}. On the other hand, some of the possible states of the system
(or more precisely, possible states of my detector interacting with
the system) can be eliminated by using some knowledge of the physics
of the system. For example, we may know some initial data or averages that characterize the
system. This becomes clear in particular if the degrees of freedom
that we choose to characterize the system with are position, momentum,
and energy, i.e., if we consider the {\em thermodynamical entropy} of
the system (see below). In this respect it is instructive to consider for a moment the continuous variable equivalent of the Shannon entropy, also known as the {\em differential entropy}, defined as~\cite{CoverThomas1991}
\be
h(X)=-\int_S f(x)\log f(x) dx\;,
\ee
with a probability density function $f(x)$ with support $S$. It turns out that while $h(X)$ is invariant with respect to translations [$h(X+c)=h(X)$], it is not invariant under arbitrary coordinate transformations: the entropy is {\em renormalized} under such changes instead. For example, 
\be
h(cX)=h(X)+\log|c|\;.
\ee
In particular, this implies that if we introduce a discretization of continuous space (e.g., via $p_i=\int_{i\Delta}^{(i+1)\Delta}f(x)dx$)
and consider the limit of the discretized version as $\Delta\to0$, we find that
\be
H_\Delta[p_i]=-\sum_ip_i\log p_i\mathop{\longrightarrow}_{\Delta\to0}h(X)-\log \Delta \;.
\ee
Thus, as the resolution of a measurement device is increased, the entropy is renormalized via an infinite term. Of course, we are used to such infinite renormalizations from quantum field theory, and just as in the field theory case, the ``unphysical" renormalization is due to an unphysical assumption about the attainable precision of measurements. Just as in quantum field theory, differences of quantities do make sense: the shared (or mutual) differential entropy is finite in the limit $\Delta\to0$ as the infinities cancel.

\subsection{Conditional Entropy}
Let us look at the basic process that reduces uncertainty: a
measurement.  When measuring the state of system $X$, I need to bring
it into contact with a system $Y$. If $Y$ is my measurement device,
then usually I can consider it to be completely known (at least, it is
completely known with respect to the degrees of freedom I {\em care}
about).  In other words, my device is in a particular state $y_0$ with
certainty. After interacting with $X$, this is not the case
anymore. Let us imagine an interaction between the systems $X$ and $Y$
that is such that
\be 
x_i y_0 \rightarrow x_iy_i\ \ \ \ \ i=1,...,N\;, \label{cmeas}
\ee 
that is, the states of the measurement device $y_i$ end up reflecting
the states of $X$.  This is a perfect measurement, since no state of
$X$ remains unresolved.  More generally, let $X$ have $N$ states while
$Y$ has $M$ states, and let us suppose that $M<N$.  Then we can
imagine that each state of $Y$ reflects an {\em average} of a number
of $X$'s states, so that the probability to find $Y$ in state $y_j$ is
given by $q_j$, where $q_j=\sum_i p_{ij}$, and $p_{ij}$ is the joint
probability to find $X$ in state $x_i$ and $Y$ in state $y_j$. The
measurement process then proceeds as
\be 
x_iy_0\rightarrow \langle x\rangle_j y_j
\ee
where 
\be 
\la x\ra_j=\sum_i p_{i|j} x_i\;. \label{cond}
\ee 
In Eq.(\ref{cond}) above, I introduced the {\em conditional
probability} 
\be
p_{i|j} = \frac{p_{ij}}{q_j}
\ee
that $X$ is in state $i$ {\em given} that $Y$ is in state
$j$. In the perfect measurement above, this probability was 1 if $i=j$
and 0 otherwise (i.e., $p_{i|j}=\delta_{ij}$), but in the imperfect
measurement, $X$ is distributed across some of its states $i$ with a
probability distribution $p_{i|j}$, for each $j$.

We can then calculate the {\em conditional entropy} (or remaining
entropy) of the system $X$ given we found $Y$ in a particular state
$y_j$ after the measurement: 
\be H(X|Y=y_j)= -\sum_i^N p_{i|j}\log p_{i|j}\;.
\ee

This remaining entropy is guaranteed to be smaller than or equal to the
unconditional entropy $H(X)$, because the worst case scenario is that
$Y$ doesn't resolve {\em any} states of $X$, in which case $p_{i|j}$ =
$p_i$. But since we didn't know anything about $X$ to begin with,
$p_i=1/N$, and thus $H(X|Y=y_j)\leq \log N$.

Let us imagine that we did learn something from the measurement of
$X$ using $Y$, and let us imagine furthermore that this knowledge is
permanent. Then we can express our new-found knowledge about $X$ by
saying that we know the probability distribution of $X$, $p_i$, and
this distribution is {\em not} the uniform distribution $p_i=1/N$. Of
course, in principle we should say that this is a conditional
probability $p_{i|j}$, but if the knowledge we have obtained is
permanent, there is no need to constantly remind ourselves that the
probability distribution is conditional on our knowledge of certain
other variables connected with $X$. We simply say that $X$ is
distributed according to $p_i$, and the entropy of $X$ is
\be
H_{\rm actual}(X)=-\sum_i \log p_i \log p_i \;. \label{shannon}
\ee

According to this strict view, all Shannon entropies of the form
(\ref{shannon}) are conditional if they are not maximal. And we can
quantify our knowledge about $X$ simply by subtracting this
uncertainty from the maximal one:
\be I= H_{\rm max}(X) - H_{\rm actual}(X)\;.\label{entdiff}
\ee
This knowledge, of course, is {\em information}. We can see from this expression that the entropy $H_{\rm max}$ can be seen as {\em potential information}: it quantifies how much is knowable about this system. If my actual entropy vanishes, then all of the potential information is realized. 

\subsection{Information}
In Eq.~(\ref{entdiff}), we quantified our knowledge about the states
of $X$ by the difference between the maximal and the actual entropy of
the system. This was a special case because we assumed that after the
measurement, $Y$ was in state $y_j$ with certainty, i.e, there is no remaining uncertainty associated with the 
measurement device $Y$ (of course, this is appropriate for a measurement device). In a more general scenario where two random variables are correlated with each other, we can imagine that $Y$ (after the interaction with $X$) instead is in
state $y_j$ with probability $q_j$ (in other words, we have reduced our uncertainty about $Y$ somewhat, but we don't know everything about it, just as for $X$).
We can then define the {\em average conditional entropy} of $X$ simply as 
\be
H(X|Y) = \sum_j q_j H(X|Y=y_j) \label{condent}
\ee
and the information that $Y$ has about $X$ is then the difference
between the unconditional entropy $H(X)$ and Eq.~(\ref{condent})
above,
\be
H(X:Y)=H(X)-H(X|Y)\;. \label{info}
\ee
The colon between $X$ and $Y$ in the expression for the information
$H(X:Y)$ is conventional, and indicates that it stands for an entropy
{\em shared} between $X$ and $Y$. According to our strict definition
of unconditional entropies given above, $H(X)=\log N$, but in the standard literature $H(X)$
refers to the actual uncertainty of $X$ given whatever knowledge
allowed me to obtain the probability distribution $p_i$, that is,
Eq.~(\ref{shannon}). In the case nothing is known {\em a priori} about $X$, Eq.~(\ref{info}) equals Eq.~(\ref{entdiff}).

Eq.~(\ref{info}) can be rewritten to display the symmetry between the
observing system and the observed:
\be
H(X:Y)=H(X)+H(Y)-H(XY)\;, \label{infosymm}
\ee
where $H(XY)$ is just the joint entropy of both $X$ and $Y$
combined. This joint entropy would equal the sum of each of $X$'s and
$Y$'s entropy only in the case that there are {\em no correlations}
between $X$'s and $Y$'s states. If that would be the case, we could not
make any predictions about $X$ just from knowing something about
$Y$. The information (\ref{infosymm}), therefore, would vanish. 

\subsubsection{Example: Thermodynamics}
We can view Thermodynamics as a particular case of Shannon
theory. First, if we agree that the degrees of freedom of interest are
position and momentum, then the maximal entropy of any system is
defined by its volume in phase space:
\be
H_{\rm max} = \log \Delta \Gamma\;, \label{thermomaxent}
\ee
where $\Delta \Gamma = \frac{\Delta p\Delta q}{k}$ is the number of
states within the phase space volume $\Delta p\Delta q$. Now the
normalization factor $k$ introduced in (\ref{thermomaxent}) clearly
serves to coarse grain the number of states, and should be
related to the resolution of our measurement device. In quantum
mechanics, of course, this factor is given by the amount of phase
space volume occupied by each quantum state, $k=(2\pi\hbar)^n$ where
$n$ is the number of degrees of freedom of the system. Does this mean
that in this case it is not my type of detector that sets the maximum
entropy of the system? Actually, this is still true, only that here we
assume a quantum mechanical perfect detector, while still averaging
over certain internal states of the system inaccessible to this detector. 

Suppose I am contemplating a system whose maximum entropy I have
determined to be Eq.~(\ref{thermomaxent}), but I have some additional
information. For example, I know that this system has been undisturbed
for a long time, and I know its total energy $E$, and perhaps even the
temperature $T$. Of course, this kind of knowledge can be obtained by
a number of different ways. It could be obtained by experiment, or it
could be obtained by inference, or theory. How does this knowledge
reduce my uncertainty? In this case, we use our {\em knowledge of physics}
to predict that the probabilities $\rho(p,q)$ going into our entropy
\be
H(p,q)=-\sum_{\Delta p\Delta q} \rho(p,q)\log \rho(p,q) \label{therment}
\ee
is given by the canonical distribution~\footnote{We set
Boltzmann's constant equal to 1 throughout. This constant, of course,
sets the scale of thermodynamical entropy, and would end up
multiplying the Shannon entropy just like any particular choice of
base for the logarithm would.}
\be
\rho(p,q)= \frac1Z e^{-E(p,q)/T}\;,
\ee
where $Z$ is the usual normalization constant, and the sum in
(\ref{therment}) goes over all positions and momenta in the phase space volume
$\Delta p\Delta q$. The amount of knowledge we have about the system according to Eq.~(\ref{entdiff}) is then just the difference between the maximal and actual uncertainties:
\be I = \log\frac{ \Delta \Gamma }Z -\frac ET\;, \label{infoclass}
\ee
where $E=\sum_{\Delta p\Delta q} \rho(p,q) E(p,q)$ is the energy of the system.

\section{Quantum Theory}
In quantum mechanics, the concept of entropy translates very easily,
but the concept of information is thorny. John von Neumann introduced
his eponymous quantum mechanical entropy as early as
1927~\cite{vonNeumann1927}, a full 21 years before Shannon introduced
its classical limit! In fact, it was von Neumann who suggested to
Shannon to call his formula (\ref{entropy}) `entropy', simply because, as he said,  ``your uncertainty
function has been used in statistical mechanics under that
name''~\cite{TribusMcIrvine1971}. 

\subsection{Measurement}
In quantum mechanics, measurement plays a prominent role, and is
still considered somewhat mysterious in many respects. The proper
theory to describe measurement dynamics in quantum physics, not
surprisingly, is quantum information theory. As in the classical
theory, the uncertainty about a quantum system can only be defined in
terms of the detector states, which in quantum mechanics is a discrete
set of {\em eigenstates} of a measurement operator. The quantum system
itself is described by a wave function, given in terms of the quantum
system's eigenbasis, which may or may not be the same as the
measurement device's basis.

For example, say we would like to ``measure an electron''. In this
case, we may mean that we would like to measure the position of an
electron, whose wave function is given by $\Psi(q)$, where $q$ is the
coordinate of the electron. Further, let the measurement 
device be characterized initially by its
eigenfunction $\phi_0(\xi)$, where $\xi$ may summarize the 
coordinates of the device. Before measurement, i.e., before the 
electron interacts with the measurement device, the system is described 
by the wave function 
\begin{equation}
\Psi(q)\phi_0(\xi)\;.
\end{equation}
After the interaction, the wave function is a superposition
of the eigenfunctions of electron and measurement device
\begin{equation}
\sum_n \psi_n(q)\phi_n(\xi)\;. \label{sum}
\end{equation}
Following orthodox measurement theory~\cite{Zurek1981}, the classical nature of the 
measurement apparatus implies that after measurement
the ``pointer'' variable $\xi$ takes on a well-defined value at each point in
time; the wave function, as it turns out, is thus {\em not} given by the
entire sum in (\ref{sum}) but rather by the single term
\begin{equation}
 \psi_n(q)\phi_n(\xi)\;.\label{term}
\end{equation}
The wave function (\ref{sum}) is said to have collapsed to (\ref{term}).
\par
Let us now study what actually happens in such a measurement in detail.
For ease of notation, let us recast this problem
into the language of state vectors instead. The first stage of the measurement 
involves the interaction of the quantum system $Q$
with the measurement device (or ``ancilla'') $A$. Both the quantum system
and the ancilla are fully determined by their state vector, yet, let us 
assume that the state of $Q$ (described by state vector $|x\rangle$)
is unknown whereas the state of 
the ancilla is prepared in a special state $|0\rangle$, say. The state 
vector of the combined system $|QA\rangle$ before measurement then is
\begin{equation}
|\Psi_{t=0}\rangle = |x\rangle|0\rangle \equiv |x,0\rangle\;.
\end{equation}
The von Neumann measurement~\cite{vonNeumann1932} is described by the unitary evolution of $QA$
via the interaction Hamiltonian
\begin{equation}
\hat H = -\hat X_Q \hat P_A \;,\label{ham}
\end{equation}
operating on the product space of $Q$ and $A$. Here, $\hat X_Q$ is the 
observable to be measured, and  
$\hat P_A$ the operator  {\it conjugate} to the
degree of freedom of $A$ that will reflect the result of the measurement.
We now obtain for the state vector $|QA\rangle$ 
after measurement (e.g., at $t=1$, putting $\hbar=1$) 
\begin{equation}
|\Psi_{t=1}\rangle=e^{i\hat X_Q \hat P_A}|x,0\rangle =
 e^{ix\hat P_A}|x,0\rangle = |x,x\rangle \;. \label{entang}
\end{equation}
Thus, the pointer variable in $A$ that previously pointed to zero now also
points to the position $x$ that $Q$ is in. This operation appears to
be very much like the classical measurement process Eq.~(\ref{cmeas}), but
it turns out to be quite different. In general, the unitary
operation~(\ref{entang}) introduces quantum {\em entanglement} between the system being measured and the measurement apparatus, a concept that is beyond the 
classical idea of correlations. 

That entanglement is very different from correlations becomes evident if we apply the unitary operation described above
to an initial quantum state which is in a quantum {\em superposition}
of two states:
\begin{equation}
|\Psi_{t=0}\rangle = |x+y,0\rangle\;.
\end{equation}
Then, the linearity of quantum mechanics implies that
\begin{equation}
|\Psi_{t=1}\rangle = e^{i\hat X_Q\hat P_M}
\biggl(|x,0\rangle+|y,0\rangle\biggr)=
|x,x\rangle+|y,y\rangle\label{entang1}\;.
\end{equation}
This state is very different from what we would expect in classical
physics, because $Q$ and $A$ are not just correlated (like, e.g., the
state $|x+y,x+y\rangle$ would be) but rather they are {\em quantum
entangled}. They now form {\em one} system that cannot be thought of
as composite. 

This nonseparability of a quantum system and the device measuring it
is at the heart of all quantum mysteries. Indeed, it is at the heart
of {\em quantum randomness}, the puzzling emergence of unpredictability
in a theory that is unitary, i.e., where all probabilities are
conserved.  What is being asked here of the measurement device, namely
to describe the system $Q$, is logically impossible because after
entanglement the system has grown to $QA$. Thus, the detector is being
asked to describe a system that is {\em larger} (with respect to the possible
number of states) than the detector, because it includes the detector {\em
itself}. This is precisely the same predicament that befalls a computer program
that is asked to determine its own halting probability, in the famous
{\it Halting Problem}~\cite{Turing1937a} analogue of G\"odel's famous Incompleteness
Theorem~\cite{Goedel1931}. Chaitin~\cite{Chaitin1997} showed that the self-referential
nature of the question that is posed to a computer program written to solve the Halting Problem gives rise to
randomness in pure Mathematics: the halting probability $\Omega=\sum_{p\ {
\rm halts}}2^{-|p|}$, where the sum goes over all the programs $p$ that halt and $|p|$ is the size of those programs, is random in every way that we measure randomness~\cite{Calude1998}. A quantum measurement is self-referential
in the same manner, since the detector is asked to describe its own
state, which is logically impossible. Thus we see that quantum
randomness has mathematical, or rather logical, randomness at its very
heart.

\subsection{von Neumann Entropy}
Because of the uncertainty inherent in standard projective measurements, measurements of a quantum system
$Q$ are described as {\it expectation values}, which are averages
of an observable over the system's {\it density matrix}, so that 
\be
\la \hat O\ra = {\rm Tr} (\rho_Q \hat O)\;, 
\ee where $\hat O$ is an operator associated with the observable we would like to measure, and 
\be
\rho_Q = {\rm Tr}_A |\Psi_{QA}\ra\la \Psi_{QA}|  \;. \label{parttrace1}
\ee  is obtained from the
quantum wave function $\Psi_{QA}$ (for the combined system $QA$, since
neither $Q$ nor the measurement device $A$ separately have a wave function after the
entanglement occurred) by {\it tracing out} the measurement device. However, technically, we are observing the states of the detector, not the states of the quantum system, so instead we need to obtain
\be
\rho_A = {\rm Tr}_Q |\Psi_{QA}\ra\la \Psi_{QA}|  \;. \label{parttrace}
\ee
by averaging over the states of the
quantum system (which strictly speaking is {\em not} being observed) and the expectation value of the measurement is instead
\be
\la \hat O\ra = {\rm Tr} (\rho_A \hat O)\;.
\ee 
The uncertainty about the quantum system is then assumed to be given by the uncertainty in the measurement device $A$, and can be calculated simply
using the von Neumann entropy~\cite{vonNeumann1927,vonNeumann1932}:
\be
S(\rho_A)= -{\rm Tr}\rho_A \log \rho_A \label{vnent}\;.
\ee

If $Q$ has been measured in $A$'s eigenbasis, then the density matrix
$\rho_A$ is diagonal, and von Neumann entropy turns into Shannon
entropy, as we expect. Indeed, measuring with respect  to the system's eigenbasis is precisely the classical limit:
entanglement does not happen under these conditions. 

\begin{figure}[htbp] %  figure placement: here, top, bottom, or page
   \centering
   \includegraphics[width=3.5in]{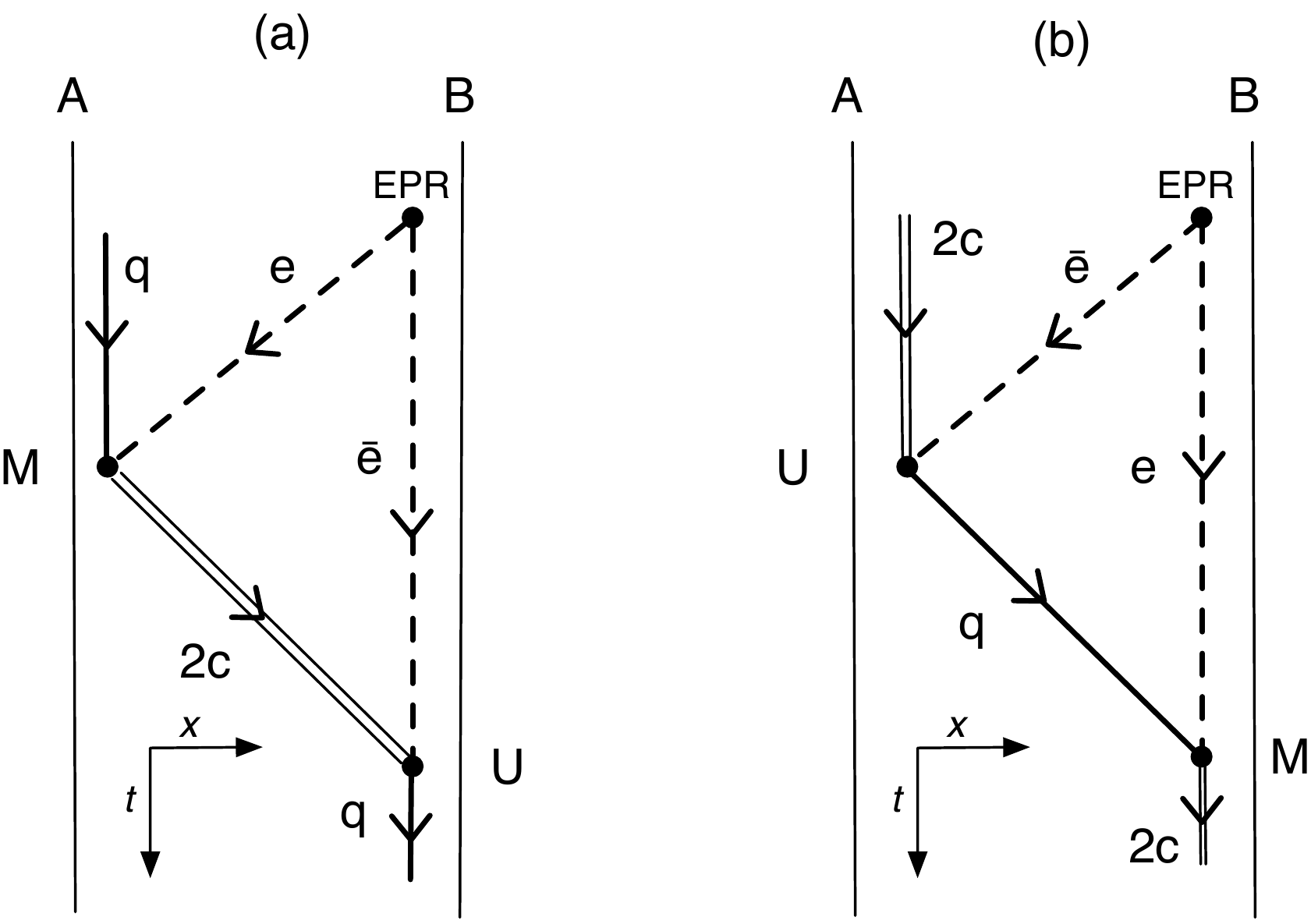} 
   \caption{Using negative partial information for quantum communication. (a) In these diagrams, time runs from top to bottom, and space is horizontal. The line marked ``A" is Alice's space-time trajectory, while the line marked ``B" is Bob's. Bob creates an $e\bar e$ pair (an Einstein-Podolski-Rosen pair) close to him, and sends the ebit over to Alice. Alice, armed with an arbitrary quantum state $q$, performs a joint measurement $M$ on both $e$ and $q$, and sends the two classical bits $2c$ she obtains from this measurement back to Bob (over a classical channel).  When Bob receives these two cbits, he performs one out of four unitary transformations $U$ on the anti-ebit he is still carrying, conditionally on the classical information he received. Having done this, he recovers the original quantum state $q$, which was ``teleported" over to him. The partial information in $e$ is one bit, while it is minus one for the anti-ebit.  (b) In superdense coding, Alice sends two classical bits of information $2c$ over to Bob, but using only a single qubit in the quantum channel. This process is in a way the ``dual" to the teleportation process, as Alice encodes the two classical bits by performing a conditional unitary operation $U$ on the anti-ebit, while it is Bob that performs the measurement $M$ on the ebit he kept and the qubit Alice sent. Figure adapted from~\protect\cite{CerfAdami1997}.}
   \label{teleport}
\end{figure}

Quantum Information Theory, of course,  needs concepts such as conditional entropies
and mutual entropies besides the von Neumann entropy. They can be defined in a straightforward
manner~\cite{Wehrl1978,CerfAdami1997}, but their interpretation needs care.  For
example, we can define a conditional entropy in analogy to Shannon
theory as 
\begarr
S(A|B)&=&S(AB)-S(B) \\
&=&-\tr_{AB}(\rho_{AB}\log \rho_{AB}) +\tr_B(\rho_B\log\rho_B)\;, \nonumber
\endarr
where $S(AB)$ is the joint entropy of two systems $A$ and $B$. But can
we write this entropy in terms of a conditional density matrix, just
as we were able to write the conditional Shannon entropy in terms of a
conditional probability? The answer is yes and no: a definition of conditional von Neumann entropy in
terms of a conditional density {\it operator} $\rho_{A|B}$
exists~\cite{CerfAdami1997,CerfAdami1999}, but this operator is technically not a
density matrix (its trace is not equal to one), and the eigenvalues of
this matrix are very peculiar: they can {\em exceed} one (this is of
course not possible for probabilities). Indeed, the eigenvalues can exceed one
only when the system is entangled. As a consequence, quantum
conditional entropies can be {\em negative}~\cite{CerfAdami1997}. This negative quantum entropy has
an operational meaning in quantum information theory: it quantifies how much additional information must be conveyed in order to transport a quantum state if part of a distributed quantum system is known~\cite{Horodeckietal2005}. If this ``partial information" is negative, the sender and receiver can use the states for future communication. In Fig.~\ref{teleport}a, we can see a quantum communication process known as 
``quantum teleportation"~\cite{Bennettetal1993}, in which the quantum wavefunction of a  {\it qubit} (the quantum analogue
of the usual bit, which is a quantum particle that can exist in
superpositions of zero and one)  is transported from the sender ``A" (often termed ``Alice") to the receiver ``B" (conventionally known as ``Bob").  This can be achieved 
using an entangled pair of particles $e\bar e$ (an ebit--anti-ebit pair), where ``ebit" stands for entangled bit~\cite{Bennettetal1996b}. This pair carries no information, but each element of the pair carries partial information: in this case the ebit carries one bit, while the anti-ebit carries minus one bit. Bob sends the ebit over to Alice, who performs a joint measurement $M$ of the pair and sends the two classical bits of information back to Bob (see Fig.~\ref{teleport}a). Armed with the two classical bits, Bob in turn can now perform a unitary operation $U$ on the anti-ebit he has been carrying around, and transform it into the original qubit that Alice had intended to convey. In this manner, Bob has used the negative ``partial information" in his anti-ebit to recover the full quantum state, using only classical information. Note that the anti-ebit with negative partial information traveling forwards in time can be seen as an ebit with positive partial information traveling backwards in time~\cite{CerfAdami1997}. The process of super-dense coding~\cite{BennettWiesner1992} can be explained in a similar manner (see Fig.~\ref{teleport}b), except here Alice manages to send 2 classical bits by encoding them on the single anti-ebit she received from Bob.

Quantum mutual entropy is perhaps even harder to understand. We can again define it simply
in analogy to (\ref{infosymm}) as~\cite{Wehrl1978,Ohya1983,CerfAdami1997}
\be S(A:B)=S(A)+S(B)-S(AB)\;, \label{mutual}
\ee
but what does it mean? For starters, this quantum mutual entropy can
be twice as large as the entropy of any of the subsystems, so $A$ and
$B$ can share more quantum entropy than they even have by themselves!
Of course, this is due to the fact, again, that ``selves'' do not
exist anymore after entanglement. Also, in the classical theory,
information, that is, shared entropy, could be used to make
predictions, and therefore to reduce the uncertainty we have about the
system that we share entropy with. But that's not possible in quantum
mechanics. If, for example, I measure the spin of a quantum particle
that is in an even superposition of its spin-up and spin-down state,
my measurement device will show me spin-up half the time, and
spin-down half the time, that is, my measurement device has an entropy
of one bit. It can also be shown that the shared entropy is {\it two
bits}~\cite{CerfAdami1997}. But this shared entropy cannot be used to
make predictions about the actual spin. Indeed, for the case of the even superposition, I still do not know
anything about it~\cite{CerfAdami1998}! On the other hand, it is possible, armed with my
measurement result, to make predictions about the state of {\em other}
detectors measuring the same spin. And even though all these detectors
will agree about their result, technically they agree about a random
variable (the state of the measurement device), not the actual state of the spin they believe their
measurement device to reflect~\cite{AdamiCerf1999}. Indeed, what else
could they agree on, since the spin does not have a state? Only the combined
system with all the measurement devices that have ever
interacted with it, does~\cite{Adami2010}. 

Still, the quantum mutual entropy plays a central role in quantum information theory, because it plays a similar role as the classical mutual entropy in the construction of the capacity of an entanglement-assisted channel~\cite{AdamiCerf1997,Bennettetal2002}. In this respect, it is unsurprising that the mutual entropy between two qubits can be as large as 2, as this is the capacity of the superdense coding channel described in Fig.~\ref{teleport}b~\cite{AdamiCerf1997}.

%Information, it turns out, is a concept that is altogether
%classical. {\it Quantum} information, in hindsight, is therefore really a contradiction in terms. But that does not mean that the entire field of quantum information theory is absurd. Rather, what we mean by ``quantum information theory'' is the study of storage, transmission, and manipulation of qubits. Indeed, the capacity of quantum channels to transmit classical information is higher than any classical channel~\cite{AdamiCerf1997,Bennettetal2002}, for example, and quantum bits can be used for super-fast computation~\cite{Shor1994}.

The extension of Shannon's theory into the quantum regime not only
throws new light on the measurement problem, but it also helps in
navigating the boundary between classical and quantum
physics. According to standard lore, quantum systems (meaning systems
described by a quantum wave function) ``become'' classical in the
macroscopic limit, that is, if the action unit associated with that
system is much larger than $\hbar$. Quantum information theory has
thoroughly refuted this notion, since we now know that macroscopic
bodies can be entangled just as microscopic ones
can~\cite{Julsgaardetal2001}. Instead, we realize that quantum systems
appear to follow the rules of classical mechanics if parts of their
wave function are averaged over [such as in Eq.~(\ref{parttrace1})],
that is, if the experimenter is not in total control of all the
degrees of freedom that make up the quantum system. Because
entanglement, once achieved, is not undone by the distance between
entangled parts, almost all systems will seem classical unless
expressly prepared, and then protected from interaction with
uncontrollable quantum systems. Unprotected quantum systems spread
their state over many variables very quickly: a process known as {\it
decoherence} of the quantum state~\cite{Zurek2003}.

\subsubsection{Quantum Thermodynamics}
A simple  example that illustrates the use of information theory in (quantum) thermal physics is the Heisenberg dimer model, defined by the Hamiltonian
\be
H=J\vec s_1\otimes \vec s_2=\frac J4 \vec \sigma \otimes\vec\sigma\;,
\ee
where $\vec \sigma=(\sigma_x,\sigma_y,\sigma_z)$ are the Pauli matrices. The system has three degenerate excited states with energy $J/4$, and a (singlet) ground state with energy $-3J/4$. 
The thermal density matrix of the two-spin system can be written in the product basis as (here, $\beta=1/T$ is the inverse temperature)
\be\label{rho}
\rho_{12}=\frac {e^{-\beta H}}Z\nonumber =
\frac{e^{\bjf}}Z\left(\begin{array}{cccc}e^{-\bjt} & 0 & 0 & 0\nonumber\\
0 & \cosh \bjt & -\sinh \bjt & 0\nonumber\\
0 & -\sinh \bjt & \cosh \bjt & 0 \nonumber \\
0 & 0 & 0 &e^{-\bjt}\end{array}\right)
\ee
where $Z={\rm Tr} e^{-\beta H}=e^{\bjth}+3e^{-\bjf}$. We can calculate the von Neumann entropy of the joint system as
\be
S(\rho_{12})=-{\rm Tr}\rho_{12}\log \rho_{12}=\log Z+\beta E\;,
\ee
where $E$ is the energy 
\be
E={\rm Tr} \rho_{12} H=\frac{3J}4\frac{1-e^{\beta J}}{3+e^{\beta J}}\;.
\ee
The marginal density matrices for each of the spin subsystems turn out to be 
\be
\rho_1=\rho_2=\left(\begin{array}{cc}1 & 0 \nonumber \\ 0 & 1\nonumber\end{array}\right)
\ee
as can easily be seen from inspecting $\rho_{12}$ above, so that $S(\rho_1)=S(\rho_2)$= 1. Using (\ref{mutual}) we can calculate the mutual entropy between the quantum subsystems to find 
\be
S(1:2)=2-\log Z -\beta E\;,
\ee 
which is formally analogous to the classical result (\ref{infoclass}), but has very peculiar quantum properties instead. In the infinite temperature limit $\beta\to 0$ we see that $Z\to 4$ while $E\to 0$, so the shared entropy vanishes in that limit as it should: no interactions can be maintained. But it is clear that at any finite temperature, the quantum interaction between the spins creates correlations that can be quantified by the mutual von Neumann entropy between the spins. In particular, in the limit of zero temperature we find
\be
\log Z +\beta E\mathop{\longrightarrow}_{T\to 0}0\;,
\ee
that is, the joint entropy of the spins $S(\rho_{12})$ vanishes and $S(1:2)\to 2$. In that case, the mutual von Neumann entropy is that of a pure Einstein-Podolski-Rosen pair: the singlet solution
\be
|\Psi\ra=\frac1{\sqrt2}\left(|\uparrow\downarrow\ra-|\downarrow\uparrow\ra\right)\;,
\ee
and exceeds by a factor of two the entropy of any of the spins it is composed of. We recognize the wavefunction of the ground state of the Heisenberg dimer at zero temperature as the entangled $e\bar e$ pair that we encountered earlier, and that was so useful in quantum teleportation and superdense coding. We will study its behavior under Lorentz transformations below. 

\section{Relativistic Theory}
Once convinced that information theory is a statistical theory about
the relative states of detectors in a physical world, it is clear that
we must worry not only about quantum detectors, but about moving ones
as well. Einstein's special relativity
established an upper limit for the speed at which information can be transmitted, without 
the need to cast this problem in an information-theoretic language. But in hindsight, it is clear that the impossibility of superluminal signaling could just as well have been the result of an analysis of the information transmission capacity of a communication channel involving detectors moving at constant speed with respect to each other. As a matter of fact, Jarett and Cover calculated the capacity of an ``additive white noise Gaussian" (AWNG) channel~\cite{CoverThomas1991} for information transmission for the case of moving observers, and found~\cite{JarettCover1981}
\be
C=W\log(1+\alpha {\rm SNR})\;, \label{awng}
\ee
where $W$ is the bandwidth of the channel, SNR is the signal-to-noise ratio, and $\alpha=\nu^\prime/\nu$ is the Doppler shift. As the relative velocity $v/c\rightarrow1$, $\alpha\rightarrow0$ and the communication capacity vanishes. In the limit $\alpha=1$, the common capacity formula for the common Gaussian channel with limited bandwidth~\cite{CoverThomas1991} is recovered. Note that in the limit of an infinite bandwidth channel, Eq.~(\ref{awng}) becomes
\be
C=\alpha\, {\rm SNR}\log_2 (e) {\rm \ bits\ per\ second}\;.
\ee

Historically, this calculation seems to have been an anomaly:
no-one else seems to have worried about an ``information theory of moving bodies'', not the least because such a
theory had, or indeed has, little immediate relevance. Interestingly, the problem that Jarett and Cover addressed with their calculation~\cite{JarettCover1981} was the famous ``twin-paradox": a thought experiment in special relativity that involves a twin journeying into space at high-speed, only to turn around to find that his identical twin that stayed behind has aged faster. Relativistic information theory gives a nice illustration of the resolution of the paradox, where the U-turn that the traveling twin must undergo creates a switch in reference frames that affects the information transmission capacities between the twins, and accounts for the differential aging.  

A standard scenario that would require relativistic information theory thus involves two random variables moving with respect to each
other. The question we may ask is whether relative motion is going to
affect any shared entropy between the variables. First, it is
important to point out that Shannon entropy is a {\em scalar}, and we
therefore do not expect it to transform under Lorentz
transformations. This is also intuitively clear if we adopt the
``strict'' interpretation of entropy as being {\em unconditional} (and
therefore equal to the logarithm of the number of degrees of
freedom). On the other hand, probability distributions (and the
associated conditional entropies) could conceivably change under
Lorentz transformations. How is this possible given the earlier
statement that entropy is a scalar? 

We can investigate this with a {\em gedankenexperiment} where the
system under consideration is an ideal gas, with particle velocities
distributed according to the Maxwell distribution. In order to define
entropies, we have to agree about  which degrees of freedom we are
interested in. Let us say that we only care about the two components
of the velocity of particles confined in the $x-y$-plane.
Even at rest, the mutual entropy between the particle velocity components $H(v_x:v_y)$ is
non-vanishing, due to the finiteness of the magnitude of $v$. A
detailed calculation~\footnote{R.M. Gingrich, unpublished} using continuous variable
entropies of the Maxwell distribution shows that, at rest
\be 
H(v_x:v_y)=\log(\pi/e) \label{bobeq}\;.
\ee
The Maxwell velocity distribution, on the other hand, will surely
change under Lorentz transformations in, say, the $x$-direction, because
clearly the components are affected differently by the boost. In
particular, it can be shown that the mutual entropy between $v_x$ and
$v_y$ will {\em rise} monotonically from $\log(\pi/e)$, and tend to a
constant value as the boost-velocity $v/c\rightarrow1$. But of course, $v/c$ is just
another variable characterizing the moving system, and if this is known
precisely, then we ought to be able to recover Eq.~(\ref{bobeq}), and
the apparent change in information is due entirely to a reduction in
the uncertainty $H(v_x)$. This example shows that in information
theory, even if the entire system's entropy does not change under
Lorentz transformations, the entropies of subsystems, and therefore
also information, can.

While a full theory of relativistic information does not exist, pieces
of such a theory can be found when digging through the literature, For
example, relativistic thermodynamics is a limiting case of
relativistic information theory, simply because as we have seen above,
thermodynamical entropy is a limiting case of Shannon entropy. But
unlike in the case constructed above, we do not have the freedom to
choose our variables in thermodynamics. Hence, the invariance of
entropy under Lorentz transformations is assured via Liouville's
theorem, because the latter guarantees that the phase-space volume
occupied by a system is invariant. Yet, relativistic thermodynamics is
an odd theory, not the least because it is intrinsically inconsistent:
the concept of equilibrium becomes dubious. In thermodynamics,
equilibrium is defined as a state where all relative motion between
the subsystems of an ensemble have ceased. Therefore, a joint system
where one part moves with a constant velocity with respect to the
other cannot be at equilibrium, and relativistic information theory
has to be used instead. 

One of the few questions of immediate relevance that relativistic
thermodynamics has been able to answer is how the temperature of an
isolated system will appear from a moving observer. Of course,
temperature itself is an equilibrium concept and therefore care must
be taken in framing this question~\cite{AldrovandiGariel1992}. Indeed, both
Einstein~\cite{Einstein1907} and Planck~\cite{Planck1908} tackled the question of how to
Lorentz-transform temperature, with different results. The
controversy~\cite{Ott1963,Arzelies1965} can be resolved by realizing that no
such transformation law can in fact exist~\cite{LandsbergMatsas1996}, as the
usual temperature (the parameter associated with the Planckian
blackbody spectrum) becomes direction-dependent if measured with a
detector moving with velocity $\beta=v/c$ and oriented at an angle
$\theta^\prime$ with respect to the radiation~\cite{Pauli1921,PeeblesWilkinson1968}
\be 
T^\prime=\frac{T\sqrt{1-\beta^2}}{1-\beta \cos\theta^\prime}\;.
\ee
In other words, an ensemble that is thermal in the rest frame is
non-thermal in a moving frame, and in particular cannot represent a
standard heat bath because it will be {\em non-isotropic}.

\section{Relativistic Quantum Theory}
While macroscopic quantities like temperature lose their meaning in
relativity, microscopic descriptions in terms of probability
distributions clearly still make sense. But in a quantum theory, these probability distributions are obtained from 
quantum measurements specified by local operators, and the space-time relationship between the detectors implementing these operators becomes important.
For example, certain measurements on a joint (i.e., composite) system may require communication between parties, while  certain others are impossible {\em even though} they do not require communication~\cite{Beckmanetal2001}. In general, 
a relativistic theory of quantum information needs to pay close attention to the behavior of the von Neumann entropy under Lorentz transformation, and how such entropies are being reduced by measurement.
In this section, I discuss the effect of a Lorentz transformation on the entropy of a single particle, or a pair of entangled particles. For the latter case, I study how quantum entanglement between particles is affected by global Lorentz boosts. This formalism has later been used to study the effect of {\em local} Lorentz transformations on the von Neumann entropy of a single particle or a pair of entangled particles, and I will summarize those results too.
  
\subsection{Boosting Quantum Entropy}
The entropy of a
qubit (which we take here for simplicity to be a
spin-1/2 particle) with wave function 
\be
|\Psi\ra=\frac1{\sqrt{|a|^2+|b|^2}}\biggl(a|\uparrow\ra+b|\downarrow\ra\biggr)
\;,\label{qubit}
\ee
($a$ and $b$ are complex numbers), 
can be written in terms of its density matrix
$\rho=|\Psi\ra\la \Psi|$ as
\be
S(\rho)=-\tr(\rho\log \rho)\;. \label{qubitent}
\ee
A wave function is by definition a completely
known state (called a ``pure state''), because the wave function is a
complete description of a quantum system. As a consequence, (\ref{qubitent}) 
vanishes: we have no uncertainty about this quantum system. 
As we have seen earlier, it is
when that wave function interacts with uncontrolled degrees of freedom
that mixed states arise. And indeed, just by Lorentz-boosting a qubit, such
mixing will arise~\cite{Peresetal2002}. The reason is not difficult to
understand.  The wave function (\ref{qubit}), even though I have just
stated that it completely describes the system, in fact only
completely describes the {\em spin} degree of freedom! Just as we saw in the
earlier discussion about the classical theory of information, there may always be
other degrees of freedom that our measurement device (here, a spin-polarization
detector) cannot resolve. Because we are dealing with particles,
ultimately we have to consider their {\it momenta}. A more
complete description of the qubit state then is 
\be 
|\Psi\ra = |\sigma\ra \times |\vec p\ra\;, \label{psqubit}
\ee 
where $\sigma$ stands for the spin-variable, and $\vec p$ is the
particle's momentum. Note that the momentum wave function $|\vec
p\ra$ is in a product state with the spin wave function
$|\sigma\ra$. This means that both spin and momentum have their own
state, they are {\it unmixed}. But as is taught in every first-year
quantum mechanics course, such momentum wave functions (plane waves
with perfectly well-defined momentum $\vec p$) do not actually exist;
in reality, they are wave packets with a momentum {\em distribution} $f(\vec
p)$, which we may take to be Gaussian. If the system is at rest, the
momentum wave function does not affect the entropy of (\ref{psqubit}),
because it is a product.

What happens if the particle is boosted? The spin and momentum degrees
{\em do} mix, which we should have expected because Lorentz
transformations always imply frame rotations as well as changes in
linear velocity. The product wave function (\ref{psqubit}) then turns into
\be
|\Psi\ra\longrightarrow \sum_\sigma \int f(\vec p) |\sigma,\vec p\ra d\vec p\;,
\ee 
which is a state where spin-degrees of freedom and momentum degrees of
freedom are entangled. But our spin-polarization detector is
insensitive to momentum! Then we have no choice but to average over
the momentum, which gives rise to a spin density matrix that is mixed
\be
\rho_\sigma={\rm Tr}_{\vec p}(|\Psi\rangle\langle\Psi |)\;,
\ee
and that consequently has positive entropy. Note, however, that the
entropy of the joint spin-momentum density matrix remains unchanged,
at zero. Note also that if the momentum of the particle was truly
perfectly known from the outset, i.e., a plane wave $|\vec p\ra$,
mixing would also not take place~\cite{AlsingMilburn2002}.

While the preceding analysis clearly shows what happens to the quantum
entropy of a spin-1/2 particle under Lorentz transformations (a
similar analysis can be done for photons~\cite{PeresTerno2003}), what is
most interesting in quantum information theory is the entanglement
{\em between} systems. While some aspects of entanglement are captured by
quantum entropies~\cite{Bennettetal1996} and the spectrum of the conditional
density operator~\cite{CerfAdami1999}, quantifying entanglement is a
surprisingly hard problem, currently without a perfect
solution. However, some good measures exist, in particular for the
entanglement between two-level systems (qubits) and three-or-fewer level systems~\cite{PlenioVirmani2007}. 

\subsection{Boosting Quantum Entanglement}
If we wish to understand what happens to the entanglement between two
massive spin-1/2 particles, say, we have to keep track of four variables, the
spin states $|\sigma\ra$ and $|\lambda\ra$ and the momentum states
$|\vec p\ra$ and $|\vec q\ra$. A Lorentz transformation on the joint
state of this two-particle system will mix spins and momenta just as
in the previous example. Let us try to find out how this affects
entanglement.

A good measure for the entanglement of mixed states, i.e., states that
are not pure such as (\ref{psqubit}), is the so-called {\it
concurrence}, introduced by Wootters~\cite{Wootters1998}. This
concurrence $C(\rho_{AB})$ can be calculated for a density matrix
$\rho_{AB}$ that describes two subsystems $A$ and $B$ of a larger
system, and quantifies the entanglement {\em between} $A$ and $B$. For
our purposes, we will be interested in the entanglement between the
spins $\sigma$ and $\lambda$ of our pair. The concurrence is one if
two degrees of freedom are perfectly entangled, and vanishes if no
entanglement is present. 

In order to do this calculation we first have
to specify our initial state. We take this to be a state with spin
and momentum wave function in a product, but where the spin-degrees of
freedom are perfectly entangled in a so-called Bell state:
\be
|\sigma,\lambda\ra = \frac1{\sqrt{2}}\left(|\uparrow,\downarrow\ra - 
|\downarrow,\uparrow\ra\right)\;. \label{bell}
\ee

The concurrence of this state can be calculated to be maximal: $C(\rho_{\sigma\lambda})=1$. We now
apply a Lorentz boost to this joint state, i.e., we move our
spin-polarization detector with speed $\beta=v/c$ with respect to this
pair (or, equivalently, we move the pair with respect to the
detector). If the momentum degrees of freedom of the particles at the
outset are Gaussian distributions unentangled with each other and the
spins, the Lorentz boost will entangle them, and the concurrence will
drop~\cite{GingrichAdami2002}. How much it drops depends on the
ratio between the spread of the momentum distribution $\sigma_r$ (not
to be confused with the spin $\sigma$) and the particle's mass $m$. In
Fig.~\ref{conc} below, the concurrence is displayed for two different
such ratios, as a function of the {\it rapidity} $\xi$. The rapidity
$\xi$ here is just a transformed velocity: $\xi = \sinh\beta$, such
that $\xi\rightarrow\infty$ as $\beta\rightarrow1$. We can see that
if the ratio $\sigma_r/m$ is not too large, the concurrence will drop but not
disappear altogether. But if the momentum spread is large compare to
the mass, all entanglement can be lost.

\begin{figure}[htb]
\centering
\includegraphics[angle=90,width=7.5cm]{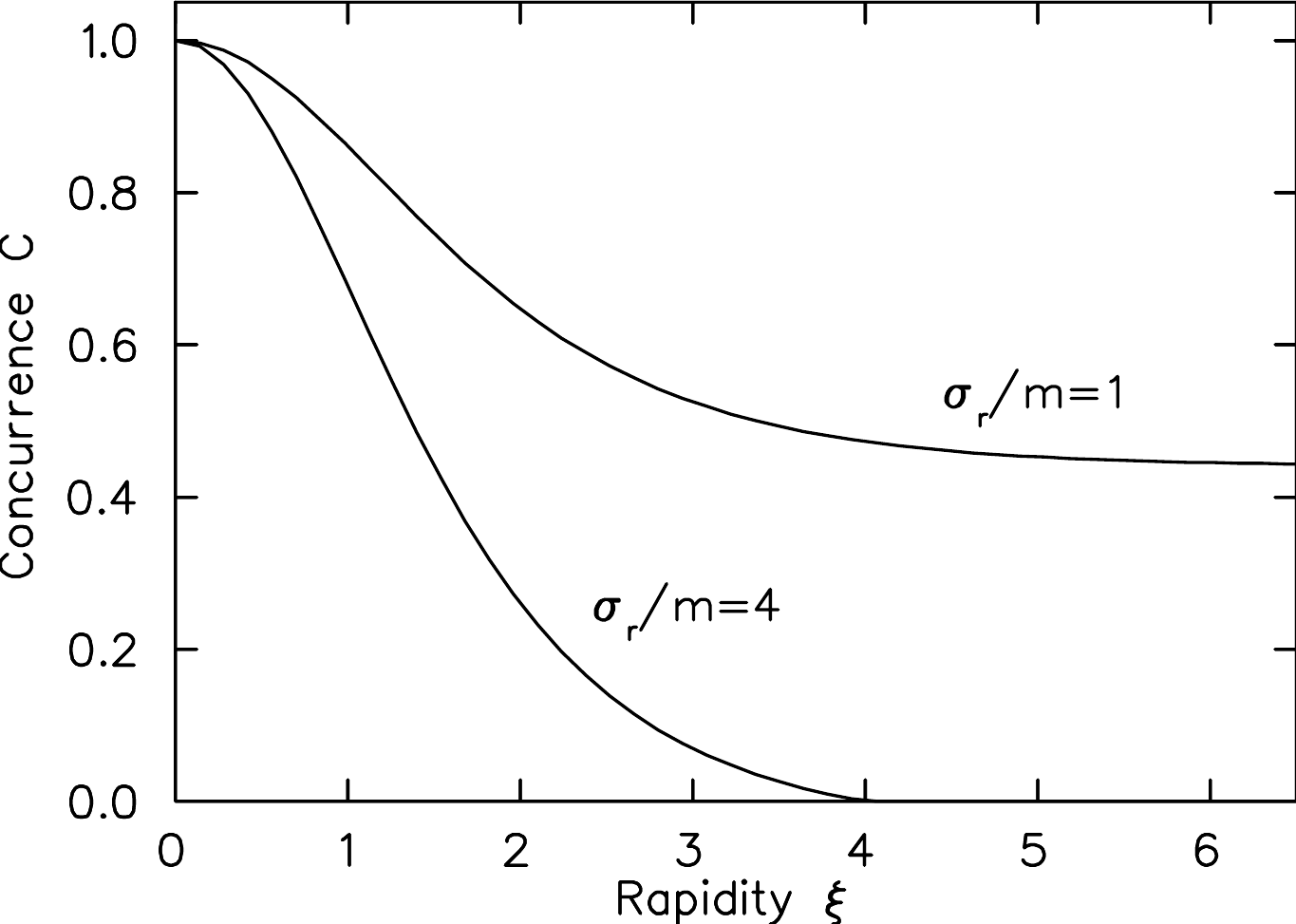}
\caption{Spin-concurrence as a function of rapidity, for an initial
Bell state with momenta in a product Gaussian.  Data is shown for
$\sigma_r / m = 1$ and $\sigma_r / m = 4$ (from
Ref.\protect\cite{GingrichAdami2002}).}.
\label{conc} 
\end{figure}

\begin{figure}[htb]
\centering
\includegraphics[angle=0,width=6.5cm]{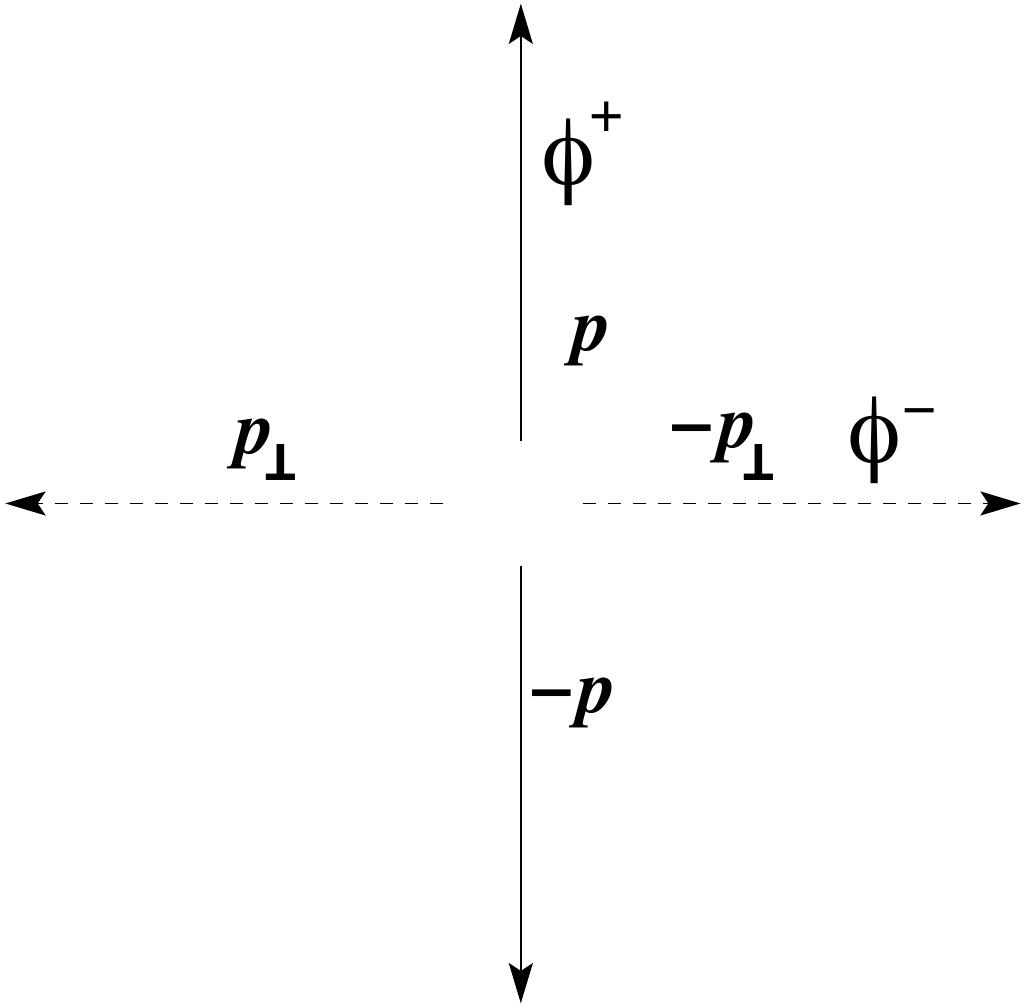}
\caption{Superposition of Bell-states $\Phi^+$ and $\Phi^-$ at right
angles, with the particle pair moving in opposite directions.}
\label{fig2} 
\end{figure}

Let us consider instead a state that is initially unentangled in spins, but
fully entangled in momenta. I depict such a wave function in
Fig.~\ref{fig2}, where a pair is in a superposition of two states, one 
moving in
opposite directions with momentum $\vec p_\bot$ in a relative spin
state $\Phi^-$ (this is one of the four Bell spin-entangled states,
Eq.~(\ref{bell})), and one moving in a plane in opposite orthogonal
directions with momentum $p$, in a relative spin-state $\Phi^+$ (which is Eq.~(\ref{bell}) but with a plus sign between the superpositions). It
can be shown that if observed at rest, the spins are indeed
unentangled. But when boosted to rapidity $\xi$, the concurrence
actually {\em increases}~\cite{GingrichAdami2002}, as for this state 
(choosing $m=1$)
\begin{equation}
C(\rho_{A B}) = \frac{p^2 (\cosh^2 (\xi) - 1)}{(\sqrt{1 + p^2} \cosh
  (\xi) + 1)^2}\;.
\end{equation}
Thus, Lorentz boosts can, under the right circumstances, create entanglement where there had been none before.

A similar analysis can be performed for pairs of entangled photons,
even though the physics is quite
different~\cite{Gingrichetal2003}. First of all, photons are
massless and their quantum degree of freedom is the
photon polarization. The masslessness of the photons makes the
analysis a bit tricky, because issues of gauge invariance enter into
the picture, and as all particles move with constant velocity (light
speed), there cannot be a spread in momentum as in the massive
case. Nevertheless, Lorentz transformation laws acting on polarization
vectors can be identified, and an analysis similar to the one
described above can be carried through. The difference is that the
entangling effect of the Lorentz boost is now entirely due to the
spread in momentum {\em direction} between the two entangled photon
beams. This implies first of all that fully-entangled photon
polarizations cannot exist, even at rest, and second that existing
entanglement can either be decreased or increased, depending on the
angle with which the pair is boosted (with respect to the angle set by
the entangled pair), and the rapidity~\cite{Gingrichetal2003}.

\subsection{Entanglement in Accelerated Frames}
The logical extension of the work just described is to allow for {\em local} Lorentz transformations on quantum particles, that is, to study particles on relativistic (accelerating) orbits or in classical gravitational fields.  Alsing and Milburn, for example, studied the quantum teleportation channel I discussed earlier~\cite{AlsingMilburn2003}. Because quantum teleportation relies on an entangled pair of particles, the fidelity of quantum teleportation (how well Bob's version of Alice's quantum state agrees with the original) would suffer if acceleration of either Bob or Alice leads to a deterioration of entanglement. This is precisely what happens, but the origin of the deterioration of entanglement is different here: it is not due to the mixing of spin and momentum degrees of freedom, but rather due to the emergence of ``Unruh radiation" in the rest frame of the accelerated observer~\cite{AlsingMilburn2003,FuentesSchulleretal2005}. Unruh radiation is a peculiar phenomenon that is due to the appearance of a sort of ``event horizon" for accelerated observers: there are regions of spacetime that are causally disconnected from an accelerated observer, and this disconnected region affects the vacuum fluctuations that occur anywhere in space~\cite{Fulling1973,Davies1975,Unruh1976}. In this sense, the Unruh radiation is analogous to Hawking radiation, which I will discuss in more detail in the following section. Unruh radiation produces thermal noise in the communication channel, which leads to the breakdown of the fidelity of quantum teleportation. Because this reasoning applies to all quantum communication that relies on the assistance of entanglement, we can conclude that generally the capacity of entanglement-assisted channels would be reduced between accelerated observers~\cite{Alsingetal2006}. A similar conclusion holds for entangled particles near strong gravitational fields~\cite{TerashimaUeda2004}. In that case, it is indeed the Hawking radiation that leads to the deterioration of entanglement between Einstein-Podolski-Rosen pairs. 

\section{Information in Curved Space Time}

While there are clearly many other questions that can conceivably be
posed (and hopefully answered) within the relatively new field of
relativistic quantum information theory~\cite{PeresTerno2004}, I would
like to close this review with some speculations about quantum
information theory in curved space time.

That something interesting might happen to entropies in curved space
time has been suspected ever since the discovery of Hawking
radiation~\cite{Hawking1975} that gave rise to the black hole information
paradox~\cite{Preskill1992}. The paradox has two parts and 
can be summarized as follows:
According to standard theory, a non-rotating and uncharged black hole can be described by an
entropy that is determined entirely in terms of its mass $M$ (in units
where $\hbar=G=1$): 
\be 
S_{BH}=4\pi M^2\;.  
\ee
Presumably, a state that is fully known (that is, one that is correlated with another system that an observer has in its possession) can be absorbed by the black hole. Once that state disappears behind the event horizon, the correlation between that state and its description in the observer's hands seems to disappear: the information
cannot be retrieved any longer. Even worse, after a long time, the
black hole will have evaporated entirely into thermal (Hawking) radiation, and
the information is not only irretrievable, it must have been destroyed. A more
technical discussion would argue that black holes appear to have the
capability to turn pure states into mixed stated {\em without}
disregarding (tracing over) parts of the wave function. Such a state of affairs is not
only paradoxical, but it is in fact completely incompatible not only
with the standard model of physics, but with the basic principle of probability conservation. 

The second part of the problem has to do with the entropy balance
between the black hole and the radiation it emits.  When the black
hole evaporates via Hawking radiation, the emitted radiation is
thermal, and carries entropy
\be
S_{\rm rad}\sim T^3_{H}
\ee
with black hole temperature $T_H=(8\pi M)^{-1}$. But the black hole's
entropy must also change at the same time, and this is determined by
the amount of energy that had to be spent in order to create the
virtual particle pairs that gave rise to the radiation. Because mass
and temperature of the black hole are inversely related,
the entropy decrease of the black hole and the entropy of the
emitted radiation {\em cannot match}. Indeed, we roughly find that
\be
d S_{\rm rad} \approx 4/3 \,\, d S_{BH}\;.
\ee

Now it should be pointed out that the preceding results were obtained
within equilibrium thermodynamics in curved space time. But since
black holes have negative heat capacity~\cite{Landsberg1988}, they can {\em never} be at
equilibrium, and the assumptions of that theory are strongly
violated. As the concept of information itself is a non-equilibrium
one, we should not be surprised if paradoxical results are obtained if equilibrium concepts are used to describe
such a case. Still, a resolution of the microscopic dynamics in black
hole evaporation is needed. One possible approach is to use quantum
information theory to characterize the relative states of the black
hole, the stimulated radiation emitted during the formation of the
black hole, and the Hawking radiation (spontaneous emission of radiation) created in the subsequent
evaporation~\cite{AdamiCerf1999,AdamiVerSteeg2004}. As we have lost track of the stimulated
radiation, we must always average over it (``trace'' it out), which (along with tracing out the causally disconnected region that lies beyond the Schwarzschild radius) creates
the positive black hole entropy. In the flat space-time
treatment of Ref.~\cite{AdamiCerf1999}, the entropy balance between the
black hole and the Hawking radiation can be maintained because
entanglement is spread between the stimulated radiation, Hawking
radiation, and the black hole. While all three are strongly entangled,
tracing over the stimulated radiation produces a state of {\em no}
correlations between Hawking radiation and black hole, implying that
the Hawking radiation appears purely thermal. But of course, the joint system is still highly entangled, but in order to discover this entanglement we would have to have access to the lost radiation emitted during the
formation process. Still, this treatment is
unsatisfying because it does not resolve the ultimate paradox: the
unitary description only works up until the black hole has shrunk to a
particular small size. At that point it appears to break down.

One reason for this breakdown might lie in the inappropriate treatment
of quantum entropy in curved space time (the preceding formalism
ignored curvature). A more thorough analysis must take into account 
the causal structure of space time. For example, not all quantum
measurements are realizable~\cite{Beckmanetal2001}, because only those
variables can be simultaneously measured whose separation is
space-like.  In physics, we do have a theory that correctly describes
how different observables interact in a manner compatible with the
causal structure of space-time, namely quantum field theory. In order
to consistently define quantum entropies then, we must define them
within quantum field theory in curved space-time.

The first steps toward such a theory involve defining quantum fields
over a manifold separated into an accessible and an inaccessible
region. This division will occur along a world-line, and we shall say
that the ``inside'' variables are accessible to me as an observer,
while the outside ones are not. Note that the inaccessibility can be
due either to causality, or due to an event horizon. Both cases can be
treated within the same formalism. States in the inaccessible region
have to be averaged over, since states that differ only in the outside
region are unresolvable. Let me denote the inside region by $R$, while
the entire state is defined on $E$. We can now define a set of
commuting variables $X$ that can be divided into $X_{\rm in}$ and
$X_{\rm out}$. By taking matrix elements of this density matrix of the
entire system
\be
\rho=|E\ra\la E|
\ee
with the complete set of variables ($X_{\rm in},X_{\rm out}$), 
we can construct the inside density matrix (defined on $R$) as
\be
\rho_{\rm in}=\tr_{X_{\rm out}}(\rho_{X_{\rm in}X_{\rm out}})\;.
\ee
which allows me to define the {\em geometric entropy}~\cite{Holzheyetal1994}
of a state $E$ for an observer restricted to $R$
\be
S_{\rm geom}= -\tr(\rho_{\rm in}\log \rho_{\rm in})\;. \label{geom}
\ee
Here, the trace is performed using the inside variables only.

This, however, is just the beginning. As with most quantities in
quantum field theory, this expression is divergent and needs to be
renormalized. Rather than being an inconvenience, this is precisely
what we should have expected: after all, we began this review by
insisting that entropies only make sense when discussed in terms of
the possible measurements that can be made to this system. This is, of
course, precisely the role of renormalization in quantum field
theory. Quantum entropies can be renormalized via a number of methods,
either using Hawking's zeta function regularization
procedure~\cite{Hawking1977} or by the ``replica trick'', writing
\be
S_{\rm geom} = -\left[\frac{d}{dn}\tr(\rho_{\rm in}^n)\right]_{n=1}\;,
\ee
and then writing $dS(n)$ in terms of the expectation value of the
stress tensor. A thorough application of this program should reveal
components of the entropy due entirely to the curvature of space-time,
and which vanish in the flat-space limit. Furthermore, the geometric
entropy can be used to write equations relating the entropy of the 
inside and the
outside space-time regions, as 
\be
S(E)=S(\rho_{\rm in, out}) = S(\rho_{\rm in}) + S(\rho_{\rm out}|\rho_{\rm in})\;.
\ee
If $S(\rho_{in})$ is the entropy of the black hole radiation (together
with the stimulated radiation), then $S(\rho_{\rm out}|\rho_{\rm in})$
is the conditional black hole entropy {\em given} the radiation field, a most
interesting quantity in black hole physics.
\section{Summary}
Entropy and information are statistical quantities describing an
observer's capability to predict the outcome of the measurement of a
physical system. Once couched in those terms, information theory
can be examined in all physically relevant limits, such as quantum,
relativistic, and gravitational. Information theory is a non-equilibrium
theory of statistical processes, and should be used under those
circumstances (such as measurement, non-equilibrium phase transitions,
etc.) where an equilibrium approach is meaningless. Because an
observer's capability to make predictions (quantified by entropy) is
not a characteristic of the object the predictions apply to, it does
not have to follow the same physical laws (such as reversibility) as
that befitting the objects. Thus, the arrow of time implied by the
loss of information under standard time-evolution is even less
mysterious than the second law of thermodynamics, which is just a
consequence of the former. 

In time, a fully relativistic theory of quantum information, defined
on curved space-time, should allow us to tackle a number of problems
in cosmology and other areas that have as yet resisted a consistent
treatment. These developments, I have no doubt, would make Shannon proud.

\vskip 1cm
\noindent{\bf Acknowledgments} I am grateful to N.J. Cerf for
years of very fruitful collaboration in quantum information theory, as
well as to R.M. Gingrich and A.J. Bergou for their joint efforts in the
relativistic theory. I would also like to acknowledge crucial
discussions on entropy, information, and black holes, with
P. Cheeseman, J.P. Dowling, and U. Yurtsever. This work was carried out
in part with support from the Army Research Office. I would like to acknowledge support from Michigan State University's 
BEACON Center for the Study of Evolution in Action, where part of this review was written.

\bibliography{Physics}
\bibliographystyle{ws-rv-van}
\end{document}